\documentclass[preprint,floatfix,aps,prb]{revtex4-1}
\usepackage{graphicx}
\usepackage{dcolumn}
\usepackage{bm}
\usepackage{multirow}
\usepackage{amsmath}
\usepackage{siunitx}
\usepackage{txfonts}
\usepackage{float}
\usepackage{url}
\usepackage{threeparttable}
\usepackage{booktabs}
\usepackage{xr}
\newcolumntype{C}{>{$}c<{$}}
\AtBeginDocument{
 \heavyrulewidth=.08em
 \lightrulewidth=.05em
 \cmidrulewidth=.03em
 \belowrulesep=.65ex
 \belowbottomsep=0pt
 \aboverulesep=.4ex
 \abovetopsep=0pt
 \cmidrulesep=\doublerulesep
 \cmidrulekern=.5em
 \defaultaddspace=.5em
}
\usepackage[version=4]{mhchem}
\usepackage{amsmath}
\makeatletter
\newcommand*{\addFileDependency}[1]{
  \typeout{(#1)}
  \@addtofilelist{#1}
  \IfFileExists{#1}{}{\typeout{No file #1.}}
}
\makeatother
 
\newcommand*{\myexternaldocument}[1]{%
    \externaldocument{#1}%
    \addFileDependency{#1.tex}%
    \addFileDependency{#1.aux}%
}
\myexternaldocument{supplemental}

\begin{document}
\title{Polarizable Potentials For Metals: The Density Readjusting Embedded Atom Method (DR-EAM)}

\author{Hemanta Bhattarai}
\affiliation{Department of Physics, \\
  University of Notre Dame, Notre Dame, Indiana 46556, USA}
\author{Kathie E. Newman}
\affiliation{Department of Physics, \\
  University of Notre Dame, Notre Dame, Indiana 46556, USA}
\author{J. Daniel Gezelter}
\email{gezelter@nd.edu}
\affiliation{251 Nieuwland Science Hall, \\
  Department of Chemistry \& Biochemistry, \\
  University of Notre Dame, Notre Dame, Indiana 46556, USA}

\date{\today}

\begin{abstract}
  In simulations of metallic interfaces, a critical aspect of metallic behavior is missing from the some of the most widely used classical molecular dynamics force fields. We present a modification of the embedded atom method (EAM) which allows for electronic polarization of the metal by treating the valence density around each atom as a fluctuating dynamical quantity. The densities are represented by a set of  additional fluctuating variables (and their conjugate momenta) which are propagated along with the nuclear coordinates. This ``density readjusting EAM'' (DR-EAM) preserves nearly all of the useful qualities of traditional EAM, including bulk elastic properties and surface energies. However, it also allows valence electron density to migrate through the metal in response to external perturbations.  We show that DR-EAM can successfully model polarization in response to external charges, capturing the image charge effect in atomistic simulations. DR-EAM also captures some of the behavior of metals in the presence of uniform electric fields, predicting surface charging and shielding internal to the metal. We further show that it predicts charge transfer between the constituent atoms in alloys, leading to novel predictions about unit cell geometries in layered L$1_0$ structures.
\end{abstract}

\maketitle

\section{Introduction}

The metallic models normally used in molecular dynamics simulations of bulk metals (EAM,\cite{Daw84,Foiles86,Johnson89,Daw89,Paradyn93,Voter95b,Lu97,Alemany98} Finnis-Sinclair,\cite{Finnis84,Sutton90} MEAM,\cite{BASKES:1994fk,Lee:2000vn,Thijsse:2002ly,Timonova:2011ve,Lee:2001qf,Beurden:2002ys} and Quantum Sutton-Chen\cite{QSC,Qi99}) have all been widely used by the materials simulation community for work on bulk and nanoparticle properties,\cite{Chui:2003fk,Wang:2005qy,Medasani:2007uq} melting,\cite{Belonoshko00,sankaranarayanan:155441,Sankaranarayanan:2005lr} fracture,\cite{Shastry:1996qg,Shastry:1998dx} crack propagation,\cite{BECQUART:1993rg} and alloying dynamics.\cite{Shibata:2002hh} One of the strengths common to all of the methods is the relatively large library of metals for which these potentials have been parameterized.\cite{Foiles86,PhysRevB.37.3924,Rifkin1992,mishin99:_inter,mishin01:cu,mishin02:b2nial,zope03:tial_ap,mishin05:phase_fe_ni,Zhou2001a,Zhou:2004b} However, none of these models allow the metal atoms to polarize, so they neglect a vital interaction with ionic or polar groups.

Streitz and Mintmire developed an electrostatic+EAM (ES+) potential
energy model for aluminum oxide surfaces,\cite{Streitz94} which
combined a variable-charge electrostatic approach with the
Finnis-Sinclair variant of EAM \cite{Finnis84}.  The ES+ potential was
originally parameterized only for Aluminum Oxide phases, but has
recently been adapted to other materials. Charge-equalization
approaches based on the extended Lagrangian charge equalization method
pioneered by Rick, Stuart, and
Berne,\cite{Rick:1994,RICK:1995ph,Rick2002} have also been utilized
to parametrize similar EAM+Charge Transfer models for Cu/\ce{CuO}
interfaces,\cite{Devine2011} titanium/\ce{TiO2}
interfaces,\cite{Ogata:1999qa} oxides of the ternary Al-Ni-Fe
alloys,\cite{Jeon2011} and binary Al-Zr oxides.\cite{Zhou:2004rw} Another charge transfer ionic-embedded atom method potential was parametrized by Zhou \textit{et al.} to study the growth of \ce{O}--\ce{Al}--\ce{Ni}--\ce{Co}--\ce{Fe} system.\cite{Zhou2005}

Our primary interest is the interactions between metals and nonreactive, but polar, species (e.g., water, carbon monoxide, ions, etc.). These interactions have a large role in the interfacial thermal conductance, the ordering of water on metal surfaces, surface restructuring under gas overpressure, surface friction, and slip/stick hydrodynamics. A charge or multipole that comes close to a conducting surface creates a disturbance in the valence electron density in the conductor. The charge density is altered only at the surface of the conductor, but the effective interactions are often treated using image charges or multipoles which are located inside the metal. To accurately capture this effect, EAM must be modified to handle perturbations in the valence electron density due to the presence of external species.  The goal is an atomistic model of a metal which can accurately reproduce the image charge effects exhibited by real metal interfaces.
Although the ES+ and related approaches that add fluctuating charges to EAM may exhibit some of these features, they have been tuned to bulk-like properties for fully or partially oxidized metals, and not for surface interactions and screening of metal interfaces.
There are two
significant assumptions made by the ES+ model:
\begin{enumerate}
\item The fluctuating charges interact primarily via Coulomb integrals
  between Slater-type orbitals centered on atomic sites.
\item The self-energy for modifying a charge on a site is essentially a parabolic function which depends on the ionization potential (IP) and electron affinity (EA) of the neutral species.
\end{enumerate}
We present here a model in which the core/valence distinction of
EAM-like models is left in place, and the fluctuating charges alter
the valence densities, which are relatively diffuse in EAM-based models.  We also parametrize the ``self'' potential as a sixth-order polynomial, using information from thermodynamically-derived models for charge transfer (\textit{i}.\textit{e}. Pauling electronegativities), experimental measures of charge mobility in metals (Hall coefficients), and experimental ionization data beyond the $+1$ and $-1$ oxidation states.

\section{Methodology}
For a collection of atoms with instantaneous positions, $\{\mathbf{r}\}$, and partial charges, $\{q\}$, the configurational potential energy in DR-EAM is given by
\begin{equation}
V\left(\{\mathbf{r}\},\{q\}\right) =  \sum_i F_i\left[ \bar{\rho}_i \right]  +\frac{1}{2} \sum_i\sum_{j \neq i} \phi_{ij}(r_{ij}, q_i, q_j)  + \frac{1}{2}\sum_i\sum_{j \neq i}  q_i q_j ~J\left(r_{ij}\right) + \sum_i V_\mathrm{self}\left(q_i\right).
\label{eq:dream}
\end{equation}
Here, $F_i \left[ \bar{\rho}_i \right]$ is an energy functional for embedding atom $i$ in a total valence density, $\bar{\rho}_i$, located at $\mathbf{r}_i$, the position of the pseudo-atom $i$ (nucleus + core electrons). $\phi_{ij}$ is a pair potential that represents the (mostly) repulsive overlap of the two pseudo-atom cores at a distance $r_{ij}=\left|\mathbf{r}_j-\mathbf{r}_i\right|$. $J\left(r_{ij}\right)$ is the Coulomb integral that accounts for electrostatic contributions from the fluctuations in the valence charge density distributions, and $V_\mathrm{self}$ is the energy for under- or over-charging each atom.

The instantaneous electron density due to the valence electrons from all the other atomic sites is computed at the location of each atom. For atom $i$, 
\begin{equation}
\bar{\rho}_i = \sum_{j\neq i} \left(1 - \frac{q_j}{N_j} \right) f_j \left(r_{ij}\right),
\label{eq:flucqdens}
\end{equation}
where $f_j\left(r\right)$ is the radial dependence of the valence density of neutral atom $j$, and $q_j$ is a dynamic charge variable that governs the instantaneous fluctuations in the valence density. $N_j$ is a ``valency count'' for atom $j$ that is determined by the number of free charge carriers in the bulk metal. Changes in the partial charge value allow for conduction band electrons $(q < 0)$ or holes $(q > 0)$ to migrate into a 
spatially localized cloud surrounding each atom.

The pair potential, $\phi_{ij}$, in DR-EAM also depends on the instantaneous valence densities at sites $i$ and $j$.
\begin{equation}\label{JohnsonModified}
    \phi_{ij}\left(r, q_i, q_j\right) = \frac{1}{2}\left[\frac{f_j(r)\left(1-\frac{q_j}{N_j}\right)}{f_i(r)\left(1-\frac{q_i}{N_i}\right)}\phi_{ii}(r)+\frac{f_i(r)\left(1-\frac{q_i}{N_i}\right)}{f_j(r)\left(1-\frac{q_j}{N_j}\right)}\phi_{jj}(r) \right],
\end{equation}
where $\phi_{ii}$ is the pair interaction of two $i$ atoms in the pure bulk metal.

A treatment of electrostatic interactions is required to account for local perturbations to the background electron density.  Here, we have adopted the damped shifted force (DSF) kernel\cite{DSF2006},
\begin{equation}
\begin{split}
J(r) =\Biggr{[}&\frac{\mathrm{erfc}\left(\alpha r\right)}{r}-\frac{\mathrm{erfc}\left(\alpha R_\mathrm{c}\right)}{R_\mathrm{c}} \\ &\left.+\left(\frac{\mathrm{erfc}\left(\alpha R_\mathrm{c}\right)}{R_\mathrm{c}^2}+\frac{2\alpha}{\pi^{1/2}}\frac{\exp\left(-\alpha^2R_\mathrm{c}^2\right)}{R_\mathrm{c}}\right)\left(r-R_\mathrm{c}\right)\ \right] \quad r\leqslant R_\textrm{c},
\label{eq:DSFPot}
\end{split}
\end{equation}
which has energies and forces that go smoothly to zero approaching a
cutoff value, $R_\textrm{c}$.  The damping parameter, $\alpha$,
describes the effective screening length of the charge, essentially treating density perturbations as Gaussians of width $\alpha^{-1}$.  Other electrostatic models, including analytical integration of the Slater Coulomb integrals, have also been adopted by other fluctuating charge approaches to polarization.\cite{Zhou2005,Rick:1994}

\subsection{The self potential}
\begin{figure}
    \centering
    \includegraphics[width=\linewidth]{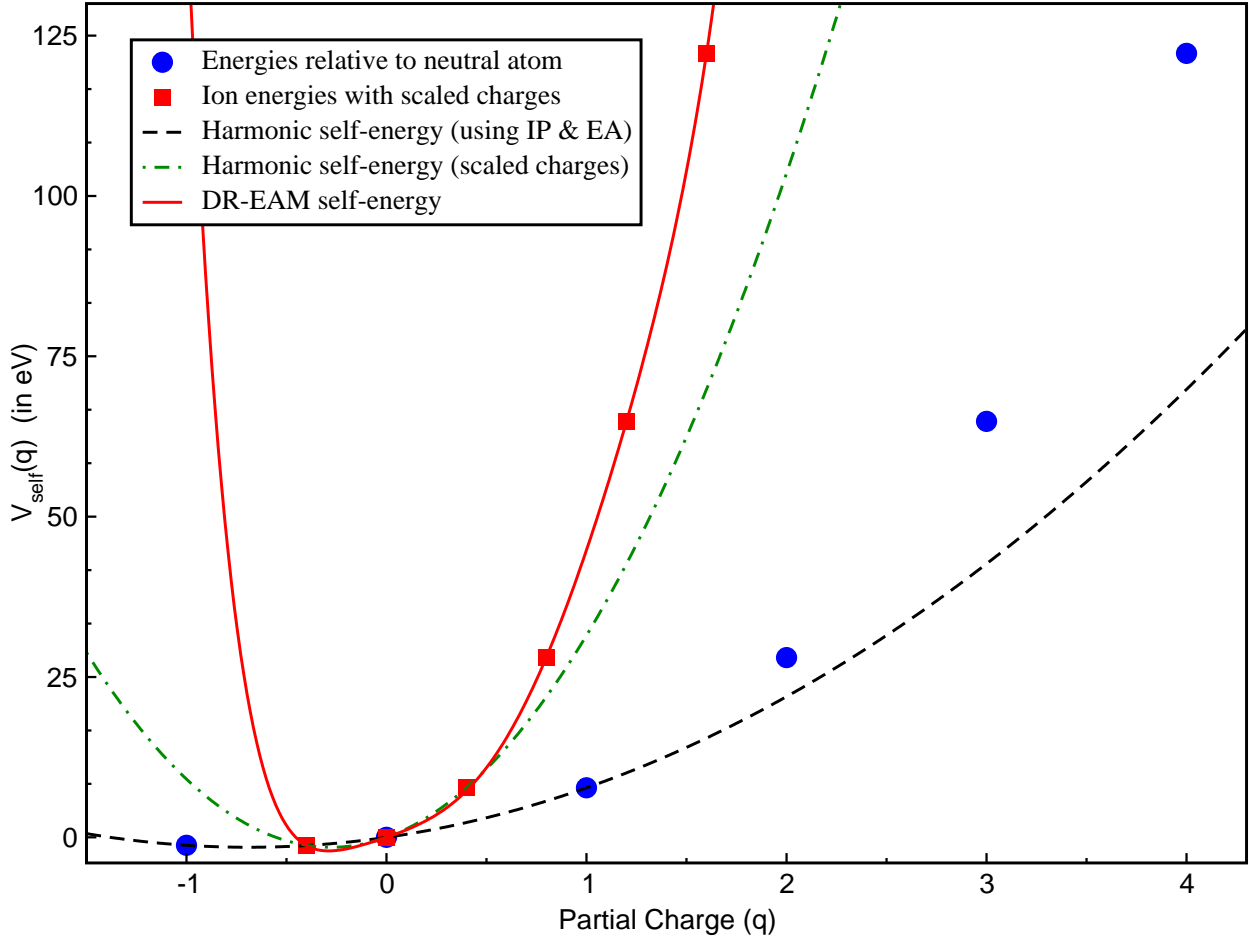}
    \caption{$V_\mathrm{self}(q)$ for Copper. In the traditional harmonic model (dashed line), the self potential is parameterized using the ionization potential (IP) and electron affinity (EA), and charge states are integer multiples of electron charge (circles). In DR-EAM, oxidation states are separated by $\sim 0.4 $e (squares), and the self potential is fit using a sixth order polynomial (solid red line).}
    \label{fig:flucq}
\end{figure}

The self potential in DR-EAM accounts for the energetic penalty for over-charging or under-charging a neutral atom, and is modeled with a polynomial,\cite{Iczkowski1961}
\begin{equation}
V_\mathrm{self}(q) =\sum_{n=1}^6 a_n~q^n .
\label{eq:self}
\end{equation}
The parameters have been tuned using a range of electron affinities and ionization potentials for commonly exhibited oxidation states in bulk materials,\cite{Hotop:1973, CuAgMo:EA, Au:EA, NiPd:EA, Co:EA, Pt:EA, Al:EA, Pb:EA, Fe:EA, Ta:EA, W:EA, Mg:EA, Ti:EA, Zr:EA} as well as the atomic polarizabilities of the neutral metals. 
See Figure. \ref{fig:flucq} for an example of the self potential in both the DR-EAM and harmonic models.

The standard electronegativity equalization models, e.g., charge equilibration (QEq)\cite{Rappe:1991os} and fluctuating charge (fluc-\textit{q})\cite{Rick:1994}, treat inter-molecular fluctuating charge interactions with a Coulombic potential, but use Slater Coulomb integrals for intra-molecular interactions, and include harmonic self potentials. The two parameters that are usually used to describe the harmonic self energy for atom $i$ are the Mulliken electronegativity,
$\chi_i$ and $J_{ii}$ (twice the chemical hardness),
\begin{equation}
\chi_i=\left(\frac{\partial V_\mathrm{self}}{\partial q}\right)_{q=0} =~ \frac{IP-EA}{2}, 
\label{eq:chi}
\end{equation}
\begin{equation}
  J_{ii} = \left(\frac{\partial^2V_\mathrm{self}}{\partial q^2}\right)_{q=0} = ~IP + EA ,
  \label{eq:J}
\end{equation} 
where energies at fixed charge states, ($q = 0,\pm 1$), i.e., the ionization potential (IP) and electron affinity (EA), are used to set both of these parameters.\cite{Rappe:1991os} However, there is now significant evidence that the charge states in the condensed phase are better represented by partial charges that are significantly smaller. 
\textit{Ab initio} calculations of ionic liquids show the charges of ions in ionic liquids is typically between 0.6 to 0.9 in the units of electron charge.\cite{Zhang:2012,Bhargava:2007,Schmidt:2010,Wendler:2011,Dommert:2012} Also, a molecular dynamics study of charge transfer in 1,3-dimethylimidazolium bis(trifluoromethylsulfonyl) imide shows the average charge on ions is $\pm$ 0.84e.\cite{Ryosuke:2016}  Pluha{\v r}ov{\'a} \textit{et al.} carried out an \textit{ab initio} molecular dynamics simulation of fluoride and lithium ion pairing in water, and found the fractional charges in fluoride (-0.79 to -0.82e) and lithium (0.89 to 0.90e).\cite{Eva:2013} 


More relevant to EAM-like models, Seriani \textit{et al.} \cite{Seriani:2007} found that in various form of Platinum Oxides, the calculated Bader charges on Pt atoms (in units of electron charge) are: 1.53 for $\alpha$-\ce{PtO2}, 1.62 for $\beta$-\ce{PtO2}, 1.13 for \ce{Pt3O4} and 0.86 for \ce{PtO} yielding an effective mapping of 0.4e for each successive oxidation state. For oxygen atoms, the charges remain roughly constant: -0.76 for $\alpha$-\ce{PtO2}, -0.81 for $\beta$-\ce{PtO2}, -0.85 for \ce{Pt3O4} and -0.86 for \ce{PtO} (all corresponding to a -2 oxidation state for oxygen).  Wang \textit{et al.} found the Bader charges in metal atoms for oxides of \ce{TiO2}, \ce{MnO2}, \ce{CoO2}, \ce{NiO2}, and \ce{ZrO2} to be 2.01e, 1.58e, 1.32e, 1.29e, 2.11e, respectively,\cite{Wang:2014} providing an average charge of 0.415e for each oxidation state. For this reason, DR-EAM adopts an effective scaling of 0.4e for the partial charge on each successive oxidation state. 

One additional consideration is the reliance on IP and EA values when other measures of electronegativity may be more useful. There are a number of approximate linear transformations between the Mulliken electronegativity and the Pauling electronegativity scale,\cite{Huheey2} which is widely used to predict polarization and charge transfer in the chemical literature. 
Because we seek to model polarization and charge transfer between a range of elements, we begin with the Pauling-Allred scale,\cite{Allred:1961fj} which is derived from thermodynamic data, and fit a linear relationship to the Mulliken scale (in eV) for all metals that have well-characterized electron affinity values. This relationship,
\begin{equation}
    \chi_\mathrm{Pauling}\approx 0.5106~\chi_\mathrm{Mulliken} - 0.2963,
\end{equation}
is then used to set the first derivative of the self potential [$a_1$ in Eq. (\ref{eq:self})]. Because we have adopted fractional partial charges (0.4e) to represent each oxidation state, the Mulliken electronegativity is related to the first-order coefficient ($a_1$) in the self potential by a factor of 2.5,
\begin{equation}\label{mulliken2a1}
a_1 = 2.5 ~ \chi_\mathrm{Mulliken} = 4.8962~\left(\chi_\mathrm{Pauling} + 0.2963\right).
\end{equation} 
For some metals (e.g., Mg), there are only approximate electron affinities available, so in order to obtain the first order coefficients in the self potential, we use tabulated Pauling electronegativity values to set $a_1$ for all of the parameterized elements. 

Traditionally, the chemical hardness has been estimated via Coulomb integrals for Slater orbitals centered on the atomic sites.\cite{Rick:1994}  Because the total charge potential in standard fluc-\textit{q} models is harmonic, the fluctuating charges then have a single low-energy solution, yielding unique sets of charges that satisfy electronegativity equalization at each atomic configuration.

In addition to overestimating of the partial charges carried by polarized sites in condensed phase simulations, we find that standard harmonic models of the self potential also overestimate the bulk polarizability (particularly for the coinage metals).
One of the parameters $(a_2)$ in the self energy (Eq. \eqref{eq:self}) can be better estimated using either empirically derived atomic polarizability data, or via a Coulomb integral of only the valence charge density.  
The chemical hardness,
\begin{equation}
    K=\frac{1}{2}\left(\frac{\partial^2 V_{\mathrm{self}}}{\partial q^2}\right)_{q=0} = \frac{1}{2}~J_{ii},
\end{equation}
has been correlated with the atomic polarizablity $\alpha$ via an empirical relationship,\cite{Nagle:1989} 
\begin{equation}\label{eqn:hardnessEmp}
 K_\textrm{empricial}=\frac{1}{4\pi\epsilon_o}\left(\frac{N}{\alpha}\right)^{\frac{1}{3}},
\end{equation}
where $N$ is the number of charge carriers per atom. The number of charge carriers per atom can be experimentally determined using the Hall effect. The Hall effect is the production of a potential difference on a current-carrying conductor when a magnetic field is applied perpendicular to the current. The Hall coefficient, $R_H$, a constant relating the magnetic field, current, and the potential produced, can be used to find the number of charge carriers.  In DR-EAM, charge carrier densities are determined using Hall coefficients ($R_H$), 
\begin{equation}\label{eqn:chargeCarrier}
 N =\frac{m}{\rho |R_{H}|}\cdot \frac{1}{N_A\mathrm{e}},
\end{equation}
where $\rho$, and $m$ are density of the metal, and the atomic mass, respectively, and $N_A$ is Avogadro's number. Each of the charge states has been scaled by 0.4, so, the number of effective charge carriers for atom $i$ is
\begin{equation}\label{eq:scaledchargecarrier}
    N_\mathrm{eff} = 0.4~\frac{m}{\rho |R_{H}|}\cdot \frac{1}{N_A\mathrm{e}}.
\end{equation}
In DR-EAM, $N_\mathrm{eff}$, is also taken to be the effective valence count in Eq.(\ref{eq:flucqdens}).

Fluctuating charge self potentials based on IP and EA can also overestimate the polarizablity of metals, sometimes badly enough to trigger a polarization catastrophe. 
This happens because the Coulombic interaction is bilinear in the charge degrees of freedom $(q_i \times q_j)$ so charges with opposing signs will tend to amplify their differences without a fast-rising self-potential to bound the charges. If the curvature of a harmonic self-potential is too small, the Coulombic interaction will dominate, leading to a polarization catastrophe.

Atomic polarizablity depends on the chemical hardness of the atom, which is related to the coefficient of second order in the self potential. In DR-EAM, the self potential is determined by imposing the constraint that the coefficient of second order is equal to the hardness of atom. The scaling of charge states and the use of a higher order polynomial helps solve the overpolarization issues. 
A fourth or sixth-order polynomial with a positive coefficient in the highest order term will eventually rise faster than the Coulombic energy falls.
The polarization catastrophe was solved by Zhou \textit{et al.} in their modified charge transfer-EAM method using upper and lower bounds on the charge values that oxygen and metals can take.\cite{Zhou:2004rw,Zhou2005}  We find that scaled charges and a higher order polynomial are sufficient to prevent the polarization catastrophe in DR-EAM.

Zhou \textit{et al.}\cite{Zhou:2004b} proposed EAM parameters and functionals for 16 metals. In this paper, we are implementing DR-EAM based on these parameters.
The valence electron density in the Zhou \textit{et al.} parametrization takes the form
\begin{equation}\label{eqn:zhoudensity}
  f(r)=\frac{f_e \exp\left[-\beta\left(\frac{r}{r_e}-1\right)\right]}{1+\left(\frac{r}{r_e}-\lambda\right)^{20}},
\end{equation}
where the denominator acts as a cutoff function that takes $f(r)$ smoothly to zero at $r / r_e$ values ranging from 2.5 to 3.5. The valence electron density without the denominator has form of a Slater function.

In the local density approximation (LDA), there is another way of estimating the chemical hardness from the electronic density of isolated atoms.\cite{Elstner:1998}
\begin{equation}
K=\frac{1}{2}~J_{ii}=\frac{1}{2}~\int\int d^3\mathbf{r}\,d^3\mathbf{r'}~ n_i(\mathbf{r})~ \frac{1}{|\mathbf{r}-\mathbf{r'}|}~ n_i(\mathbf{r'}),
\end{equation} 
where $n_i(\textbf{r})$ is the electronic density of the isolated atom.
Elstner \textit{et al.}\cite{Elstner:1998} used normalized spherical charge densities,
\begin{equation}\label{eqn:slaterform}
  n(\mathbf{r})=\frac{\tau_\alpha^3}{8\pi}e^{-\tau_\alpha|\mathbf{r}-\mathbf{R}|},
\end{equation}
where $\mathbf{R}$ is the location of the atom, to derive the chemical hardness of a spin-unpolarized atom or Hubbard parameters $U_\alpha $ in terms of $\tau_\alpha$.
\begin{equation}\label{hubbard}
  U_\alpha = \frac{1}{4\pi\epsilon_0} \frac{5}{16} \tau_\alpha ,
\end{equation}

We have adopted the Elstner \textit{et al.} result along with the Slater-form of the EAM densities in Eq. (\ref{eqn:zhoudensity}).  We find that by setting $\tau_\alpha= \beta / r_e$, a chemical hardness appropriate for use the self potential can come directly from the valence density functions,
\begin{equation}\label{eqn:hardnessTheo}
 K_\mathrm{Slater} = \frac{1}{4\pi\epsilon_0} \frac{5}{16} \frac{\beta} {r_e}\,.
\end{equation}
The valency count, $N_\mathrm{eff}$, and the hardness, $K$, values that were used to parametrize DR-EAM are provided in Table \ref{tab:KTable}.

\begin{table}
\centering
\caption{DR-EAM uses a count of the effective number of charge carriers per atom ($N_\mathrm{eff}$) as a valency count in Eq. (\ref{eq:flucqdens}) and the chemical hardness, $K_\mathrm{Slater}$, defined in Eq. (\ref{eqn:hardnessTheo}), to parametrize the harmonic term $(a_2)$ in the self potential. NB: $R_H$ is typically reported in units of $\mathrm{cm}^3 \mathrm{~C}^{-1}$ \label{tab:KTable}} 
\begin{threeparttable}
\begin{tabular}{l|cccc} 
\toprule %
&\multirow{2}{*}{$\frac{m}{\rho |R_{H}|}\cdot \frac{1}{N_A\mathrm{e}}$}  &\multirow{2}{*}{N$_\mathrm{eff}$}&\multicolumn{2}{c}{K}\\ \cline{4-5}
 & & & $K_\mathrm{Slater}$ & $K_\mathrm{empirical}$\\
\midrule
Cu & 1.42 & 0.57 & 7.63 & 7.87\\
Ag & 1.21 & 0.48 & 7.58 & 8.10\\
Au & 1.48 & 0.59 & 7.92 & 9.38\\
Ni & 1.12 &	0.45 & 8.08 & 7.78\\
Pd & 1.21 &	0.48 & 7.59 & 9.13\\
Pt & 4.10 & 1.64 & 6.15 & 12.34\\
Al & 3.05 &	1.22 & 5.54 & 11.01\\
Pb\tnote{1} & 4.57 & 1.83 & 6.70 & 12.50\\
Fe & 3.00 &	1.20 & 9.49 & 9.88\\
Mo & 0.77 &	0.31 & 7.38 & 5.62\\
Ta & 1.22 &	0.49 & 7.12 & 7.51\\
W  & 0.86 & 0.34 & 7.79 & 6.33\\
Mg & 1.68 &	0.67 & 7.68 & 8.31\\
Co & 0.82 &	0.33 & 8.31 & 6.86\\
Ti\tnote{2} & 1.15 & 0.46 & 7.18 & 7.15\\
Zr & 1.12 & 0.45 & 6.24 & 5.85\\
\bottomrule
\end{tabular}
\begin{tablenotes}
\item[1] For \ce{Pb}, we use the density and Hall coefficient of the liquid phase.
\item[2] The Hall coefficient used for \ce{Ti} is the $\frac{2}{3}$ of the perpendicular and $\frac{1}{3}$ of the parallel components.
\end{tablenotes}
\end{threeparttable}
\end{table}

\begin{figure}[h]
\centering
  \includegraphics[width=0.95\linewidth]{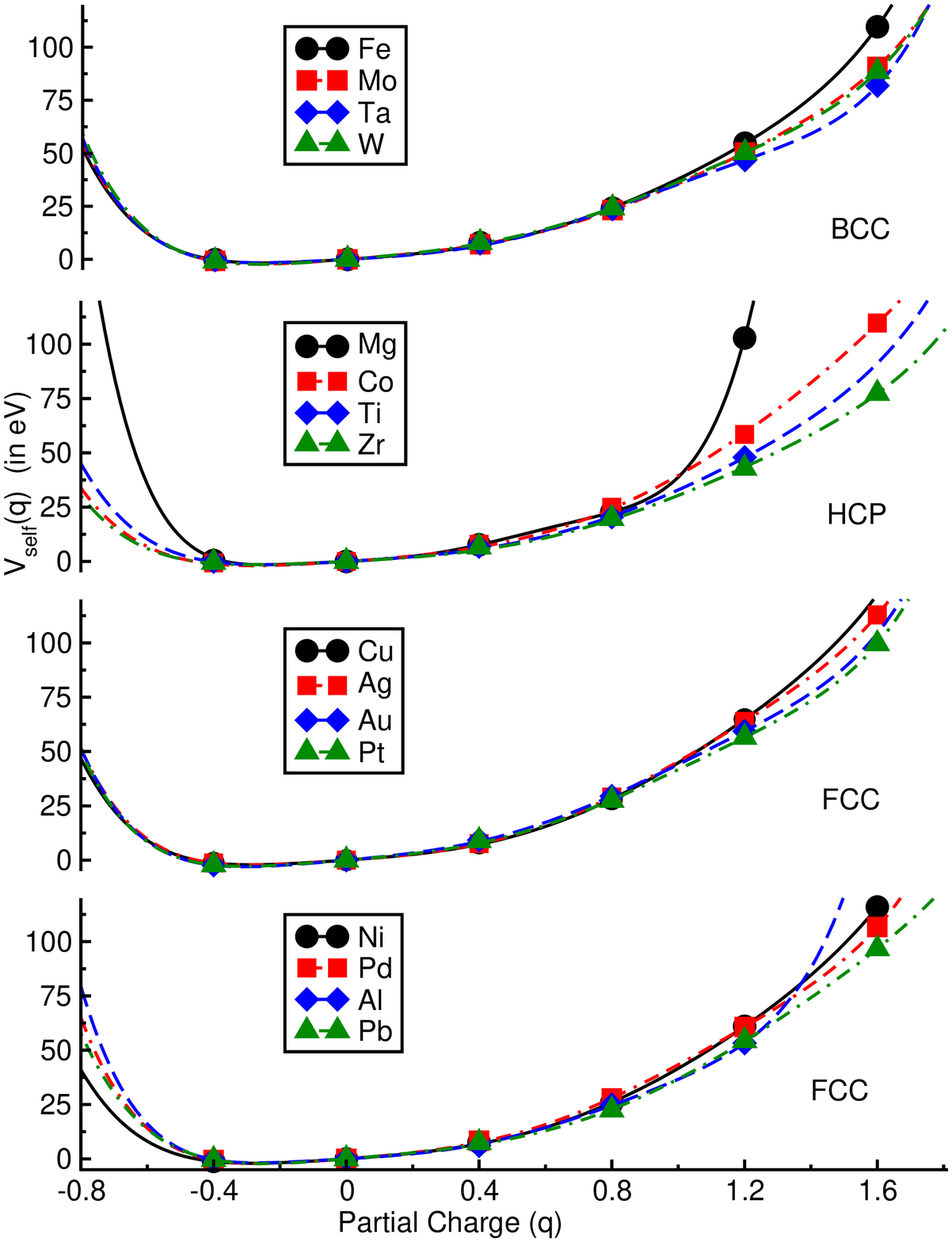}
  
\caption{DR-EAM self potential V$_\mathrm{self}(q)$ for FCC, BCC and HCP metals. The symbols are gas phase ion energies referenced to the neutral atom (with charges scaled by 0.4). Lines are Eq. (\ref{eq:self}) with coefficients given in Table. \ref{table:selfFit}. The region between -0.4 $<$ $q$ $<$ 0.4 is enlarged in the Supplemental Material\cite{Supplemental} in Fig. S1.}
\label{fig:self}
\end{figure}

The second-order coefficient ($a_2$) is fixed to value of chemical hardness ($K$) obtained from Eq. (\ref{eqn:hardnessTheo}). All remaining coefficients are determined by fitting gas phase electron affinities and ionization potentials where the charges have been scaled to match the 0.4e per oxidation state in the condensed phase. While performing the fits, the coefficients of the highest even order in the polynomial (typically $q^6$) were also constrained to be positive.  
Self-potential parameters are given in Table \ref{table:selfFit}, and plots of the DR-EAM self-potentials are given in Figure. \ref{fig:self}.

\begin{table}
\caption{Coefficients for the self potential [Eq. \eqref{eq:self}] used in DR-EAM. All parameters have units that give the self potential in eV. \label{table:selfFit}}
\centering
\begin{tabular}{c r r r r r r}
\toprule
Element & $a_1$ & $a_2$ & $a_3$ & $a_4$ & $a_5$ & $a_6$\\
\midrule
Cu&10.75& 7.63&  11.12& 65.63&  -72.22& 21.88\\      
Ag&10.90& 7.58&  14.04& 70.03&  -82.35& 25.21\\
Au&13.89& 7.92&  13.19& 68.17&  -88.02& 28.85\\
Ni&10.80& 8.08&  5.62 & 58.45&  -58.13& 16.77\\
Pd&12.22& 7.59&  2.36 & 84.42&  -89.92& 26.68\\
Pt&12.61& 6.15&  14.07& 67.32&  -86.75& 28.28\\
Al&9.33 & 5.54&  19.18& 81.65&  -136.26& 57.17\\
Pb&12.86& 6.70&  -16.03 & 79.12&  -59.70& 13.84\\
Fe&10.41& 9.49&  1.14 & 64.45&  -69.63& 22.15\\
Mo&12.30& 7.38&  -5.77& 70.98&  -68.99& 19.90\\
Ta&8.80 & 7.12&  13.48& 67.67&  -92.80& 31.22\\
W &13.01& 7.79&  -4.39& 74.83&  -78.12& 23.51\\
Mg&7.86 & 7.68&  52.67& 97.79&  -338.60& 211.34\\
Co&10.66& 8.31&  0.67 & 51.63&  -41.62& 9.92\\
Ti&8.99 & 7.18&  -1.57& 58.59&  -56.70& 16.41\\
Zr&7.96 & 6.42&  5.11 & 42.34&  -44.13& 12.65\\
\bottomrule
\end{tabular}
\end{table}

\subsection{Differences from standard EAM}
The embedding functional and pair potential are not unique in EAM, and it is possible to parametrize reasonable metal potentials by transferring some weight between the embedding functional and the effective pair potential. In DR-EAM, we have adopted the same form and parametrization for the embedding functional in DR-EAM adopted by Zhou \textit{et al.}\cite{Zhou2001a,Wadley2001a,Zhou:2004b}. Although the original EAM potentials have only positional degrees of freedom that alter the local valence density, DR-EAM adds an additional charge degree of freedom to each atomic site. The system evolves with the time-dependent positions and partial charges, readjusting the valence density contributed by each atom. The adjusted valence density in DR-EAM,
\begin{equation}
    f_j(r, q_j) \rightarrow \left(1-\frac{q_j}{N_j}\right)f_j(r),
\end{equation}
where $q_j$ and $N_j$ are the partial charge and valency count for atom $j$.

The form of the pair potential is tied to the choice of embedding functional. It is possible to transform $F$ and $\phi$ so that the slope of the embedding function is zero at the equilibrium electron density.
Mixed-element pair potentials $(\phi_{ij})$ can be defined many ways in terms of pair potentials for the two elements separately ($\phi_{ii}$ and $\phi_{jj}$).  Arithmetic means and geometric means have both been used previously,\cite{Daw89} but when the slope of the embedding function is set to zero at the equilibrium electron density, the effective two body potentials $(\phi)$ are negative at some distances, and the geometric mean cannot be used.
Johnson derived a mixed-element pair potential for alloys in terms of the valence densities,\cite{Johnson89}
\begin{equation}
    \phi_{ij}(r)=\frac{1}{2}\left[\frac{f_j(r)}{f_i(r)}\phi_{ii}(r)+\frac{f_i(r)}{f_j(r)}\phi_{jj}(r) \right].
    \label{eq:johnsonMix}
\end{equation}
This form of the pair potential is independent of the arbitrary electron-density transformation.
Since DR-EAM uses the readjusted valence density, the pair potential in DR-EAM is given by Eq. (\ref{JohnsonModified}).
The self energy, V$_\mathrm{self}$, and the Coulombic interactions could also be included in a modified functional, but for simplicity, DR-EAM treats the energies due to alteration of net atomic charges as separate terms in the potential energy.

\subsection{Charge conservation}
Because the fluctuating charge variables represent physical charge densities, they are not independent variables, and a charge conservation constraint is required when propagating these degrees of freedom along with the nuclear coordinates. Charge conservation is implemented using the method of undetermined multipliers to enforce the constraints. The net charge constraint is written as
\begin{equation}
    g(\left\{q\right\}) = \sum_i^N q_i - Q ,
\end{equation}
where $Q$ is the (fixed) charge on the system in units of electron charge and $q_i$ is the fluctuating charge of the $\mathrm{i^{th}}$ atom. Each contiguous metallic region can be thought of as a single molecule, so it is also possible to constrain metallic regions separately using several undetermined multipliers. However, using multiple constraints prevents charge transfer between the separate blocks.

\subsection{Dynamics}
The minimum energy of the fluctuating charge system is first determined using steepest-descent minimization on the configurational energy of the system with frozen nuclear coordinates. Once an optimal set of charges is known, perturbations to these charges depend only on nuclear coordinates (which are relatively slowly moving degrees
of freedom). For this reason, propagation of the fluctuating charge
variables using extended-Lagrangian simulations has become widespread.

The extended Lagrangian in DR-EAM (with the charge constraint) is given by
\begin{equation}\label{eq:extLag}
\mathcal{L}=\sum_{i=1}^N\left[\frac{1}{2}m_i \dot{\mathbf{r}}_i^2+\frac{1}{2}M_q \dot{q}_i^2\right]-V\left( \left\{ \mathbf{r} \right\}, \left\{ q \right\} \right) -\lambda \left(\sum_{i=1}^Nq_i-Q\right),
\end{equation}
where $m_i$ is the mass of the $i^\mathrm{th}$ atom, $M_q$ is the fictitious mass assigned to the fluctuating charges and $\lambda$ is Lagrange multiplier for the charge constraint. The first term in the kinetic energy is the contribution from the motion of atoms themselves while the second term is a fictitious kinetic energy, the contribution from the charge velocities $\left\{\dot{q}\right\}$.  $V\left( \left\{ \mathbf{r} \right\}, \left\{ q \right\} \right)$ is the configurational energy for the extended system defined in Eq. (\ref{eq:dream}).

Equations of motion for the fictitious charge $q$ and nuclear coordinates $\mathbf{r}$ can be obtained by calculating the forces on the dynamical charge variables and on the atoms themselves, 
\begin{equation}
    m_i ~\Ddot{\mathbf{r}}_i = \textbf{F}_i = \frac{\partial \mathcal{L}}{\partial \textbf{r}_i} = - \frac{\partial V}{\partial \mathbf{r}_i} ~,
    \label{eq:forceNuclear}
\end{equation}
\begin{equation}
    M_q ~\Ddot{q}_i = f_i = \frac{\partial \mathcal{L}}{\partial q_i} = -\frac{\partial V}{\partial q_i} - \lambda ~,
    \label{eq:forceFQ}
\end{equation}
where $\lambda$ is the charge conservation constraint,
\begin{equation}
    \lambda = - \frac{1}{N} \sum_{i=1}^{N} \frac{\partial V}{\partial q_i}~.
    \label{eq:constraint}
\end{equation}

 Two temperatures, nuclear and electronic, are defined for the system. The nuclear temperature arises from the kinetic energy of atoms whereas the electronic temperature comes from the fictitious kinetic energy of the fluctuating charges.  
 Perturbations in the fluctuating charge forces will be due primarily to the motion of surface-adsorbed molecules (e.g., water), so the timescale for charge fluctuations should be approximately the same as for nuclear motion.
 For this reason, $M_q$ is chosen to be large enough ($M_q = 600 \mathrm{~kcal~mol^{-1}~fs^2~e^{-2}}$) so that changes in electronic degrees of freedom can be integrated along with nuclear coordinates. 
 In theory, if $M_q$ is chosen to be very small, it would be possible to simulate the collective electronic motion (e.g., plasmons), but the time steps for those simulations would by necessity be extremely small relative to molecular motion.

The charge and nuclear degrees of freedom are coupled by the DR-EAM potential energy in Eq. (\ref{eq:dream}).  Although it is possible to propagate the entire system in the microcanonical (NVE) ensemble, equipartition eventually brings the electronic temperature to the same value as the nuclear temperature.  Electronegativity equalization methods assume a \textit{single} low-energy solution to the charge degrees of freedom, i.e., an electronic temperature of zero, and there are many ways to achieve this.  One could minimize the energy with respect to the charge degrees of freedom at every time step, but this would be prohibitively expensive.  It is also possible to propagate the nuclear coordinates in the microcanonical (NVE) ensemble while simultaneously keeping the electronic degrees of freedom at a much lower temperature, $T_e < 1$~K,  with a Nos\'{e}-Hoover thermostat. In many of the tests and simulations described below, the electronic degrees of freedom were kept at a low temperature using Langevin dynamics, with a drag coefficient of $0.1 \mathrm{~kcal~mol^{-1}~fs~e^{-2}}$.
As in any fluctuating charge method, the coupling between the electronic and nuclear degrees of freedom will eventually transfer energy between the subsystems, so we expect that separately thermostatting the electronic coordinates will be required.

We note that the fluctuating charge forces modified by the charge conservation constraint are used in all steps of this method, including the initial minimization to find the optimal set of charges.   

\section{Tests and Applications}
To test the new method, we carried out simulations using DR-EAM where the choice of the underlying EAM functions and parameters were taken from Zhou \textit{et al.}\cite{Zhou:2004b} as a basis for further refinement. We have tested the new method on: (1) pure bulk metals, (2) common surfaces of the bulk metals, (3) ordered structures (L$1_0$ and L$1_2$), (4) a metal slab in a uniform electric field, and (4) metal slabs with external fixed charges approaching the surface. 
In all tests, electrostatic interactions were calculated with the damped shifted force (DSF) method with a damping parameter ($\alpha$) of 0.14~\si{\angstrom}$^{-1}$.

\subsection{Bulk metals}
The cohesive energy per atom, vacancy formation energy, bulk modulus, shear modulus, and Poisson's ratio were computed for both DR-EAM and the unmodified EAM energy function,
and these are provided in Table \ref{tab:Mechanical}.
Elastic stiffness tensors were calculated using the Energy versus Strain method of Yu \textit{et al.}\cite{Yu:2010qm,Golesorkhtabar:2013cr} and elastic constants were computed using the same definitions that are utilized by the Materials Project.\cite{Jong:2015bs,Gaillac:2016eu}  

In all of the parameterized elements, the bulk metals do not polarize to any significant degree. There is negligible (nearly zero) charge transfer between the atoms in pure metals, so DR-EAM and EAM produce identical bulk properties within the limit of precision used in calculation. 

\begin{turnpage}

\begin{table}

\caption{Comparison of Mechanical Properties for Bulk Metals. For each element, the top row includes the cohesive energy per atom ($E_\mathrm{coh}$), vacancy formation energy ($E_v^f$), bulk modulus ($B$), shear modulus ($G$), and Poisson's ratio ($\sigma$).  These values (identical for DR-EAM and the unmodified EAM model within numerical precision) are compared with a second line from DFT calculations (when available).  A third line with experimental values is also presented.\cite{McKee:1972,Medasani:2015,mpApi,FeVacancy,MgVacancy,ZrCohesive,Kittel8,Simmons1971}} 
\centering 

\setlength{\tabcolsep}{6pt}
\def\arraystretch{0.55}

\begin{tabular}{ r| r r r r r r |r|r r r r r r} 
\toprule 
&E$_\mathrm{coh}$ (eV) &E$_v^f$ (eV) & B (GPa)& G (GPa) &$\sigma$& &  &E$_\mathrm{coh}$ (eV) &E$_v^f$ (eV) & B (GPa)& G (GPa) &$\sigma$
\\
\midrule 
Cu	& 3.65&	1.42&	125.74&	64.48&	0.28     && Al	&3.74&	0.74&	52.76&	17.52&	0.28\\
	& 4.09&	-	& 145	& 57	& 0.34       && 	& 3.74	&	-	& 83 &25	 &0.37\\
	& 3.43 & 1.28 & 140 & 59.3      & 0.31           &&     & 3.39  & 0.68  & 76 & 29.0      &0.35 \\
\midrule
Ag	& 2.89&	1.12&	105.77&	41.29&	0.32 && Pb	&2.01&	0.60&	47.65&	13.10&	0.34\\
	& 2.83	&-	&	88	&28	&0.38 &&            & 3.71	&-	&	37	&14	&0.35\\
	& 2.95 & 1.11 &100 & 37.5& 0.34       &&            & 2.03 & 0.50 & 46 &15.7  &0.37\\
\midrule
Au	& 3.94  &	0.92&	169.65&	38.39&	0.39 && Fe	&4.34&	1.65&	117.78&	58.08&	0.39\\
	& 3.27	&	-	&   137		&19	 &0.45   &&     & 8.45	& -  &	182	  &78	 &0.32\\
	& 3.81  & 0.89  &180        & 33.7   & 0.41      &&     & 4.28  & 1.40& 170   &94.1  &0.27\\
\midrule
Ni	&4.51&	1.78&	146.23&	74.41&	0.28    &&  Mo	&7.07&	2.76&	179.48&	60.70&	0.31\\
	& 5.78	&	-	& 198		&102	&0.29&&     & 10.86	&	-	& 262		&127	&0.30\\
	& 4.44 & 1.79&180& 101.1&0.27                    &&     & 6.82 & 3.0 & 230 &130.4 &0.28 \\
\midrule
Pd	&4.00&	1.56&	190.36&	62.04&	0.35 && Ta	&8.37&	2.80&	170.83&	39.06&	0.28\\
	& 5.18	&	-	&	160	&50	&0.38    &&       & 11.85	&-	&  194		&63	&0.35\\
	& 3.89 & 1.85 &180  &54.3 &0.37    &&                & 8.10 & -& 200 &74.0  &0.33\\
\midrule
Pt	&5.98&	1.34&	252.62&	56.49&	0.37 && W	&8.95&	3.62&	191.97&	82.61&	0.31\\
	& 6.05	&-	& 247		&49	&0.41    &&     & 12.95	&-	& 304		&147	&0.29\\
	& 5.84 & 1.32 & 230 &67.3 & 0.39           &&     & 8.90 & 4.0 & 310 &163.4  &0.27\\
\midrule
Mg	&1.54&	0.65&	34.83&	17.19&	0.28   && Co	&4.40&	1.83&	190.54&	79.99&	0.31\\
	& 1.59	&	-	&	37	&18	&0.29      &&       & 7.11	&	-	& 212 & 106 & 0.29\\
	& 1.51 & 0.79 & 35.4 &19.4  &0.27             &&       & 4.39 & 1.34 & 191.4 &91.1 &0.29\\

\midrule
Ti	&4.54&	0.77&	214.06&	109.5&	0.28 && Zr	&6.33&	1.95&	89.28&	31.92&	0.34\\
	& 7.88	&	-	& 113		&47	&0.32 && 	& 8.54	&	-	&	94	&35	&0.34\\
	& 4.85&1.55 & 110 &50.5 &0.30  && 	                & 6.29& - & 83.3 & 42.5& 0.31\\

\bottomrule
\end{tabular}
\label{tab:Mechanical}

\end{table}
\end{turnpage}
\subsection{Surface energies of bulk metals}
In DR-EAM, the anisotropic environment around atoms at the surface of a metal encourages charge transfer into the near-surface atoms to allow the surface atoms to approach the equilibrium valence density ($\rho_e$). In traditional EAM, the only way for atoms to approach equilibrium densities is by contracting the surface layers closer to the underlying bulk.

To study this difference, the surface formation energies for (111), (110), and (100) surfaces are computed for FCC and BCC metals using both DR-EAM and EAM. For HCP metals, the basal plane (0001) was exposed to calculate the surface energy. 
This data is provided in Table \ref{tab:SurfaceEnergies}.
For most of the surfaces, DR-EAM and EAM again produce identical surface energies within the limit of precision used in the calculations. Notable exceptions are the Ag(110), Al(100), and Al(110) surfaces. Even for these surfaces, however, the surface polarization yields only a small charge, $\langle q \rangle < 0.0025$, on the atoms in the subsurface layers.  Surface atoms exhibit even smaller charges.

\begin{turnpage}
\begin{table}

\caption{Comparison of surface energies (all reported in J m$^{-2}$), and average atomic charge on the surface layers $\langle q \rangle$ (in e).  Experimental surface energies are averaged over many exposed facets.\cite{Vitos:1998} In the first data column, the DR-EAM surface energy is provided. When nonpolarizable EAM predicts a different surface energy, this value is given in parentheses. Values of $\langle q \rangle$ smaller than $10^{-4}$e are indicated with an asterisk.} 
\centering 
\setlength{\tabcolsep}{6pt}
\def\arraystretch{0.55}
\begin{tabular}{r| c c c c c c | r| c c c c c} 
\toprule
    &Surface    &DR-EAM (EAM) &DFT~\cite{mpApi} &Experiment~\cite{Vitos:1998}& $\langle q \rangle$  & & &Surface &DR-EAM (EAM)&DFT~\cite{mpApi} &Experiment~\cite{Vitos:1998} & $\langle q \rangle$\\
\midrule
	& (111)     &1.599  & 1.31  &         & * &&       & (111)&    0.352 & 0.25    &          & *          \\
Cu	& (100)     &1.655  & 1.47  &  1.790  & * &&   Pb  & (100)&    0.395  & 0.28    &   0.593  &	*       \\
	& (110)     &1.852  & 1.56  &	      & * &&       & (110)&    0.428 & 0.33    & 	      & *            \\
	\midrule 
	& (111)     & 0.947         & 0.77 &           & *      &&          & (111) & 1.988 & 2.73  &          & * \\
Ag	& (100)     & 1.025         & 0.81 &    1.246  & *      &&       Fe & (100) & 1.441 & 2.50  &   2.417  & *	\\
	& (110)     & 1.119 (1.115) & 0.87 &	       &  0.001 &&          & (110) & 1.350 & 2.45  & 	       & * \\
\midrule 
	& (111) & 0.923 & 0.74  &           & * &&              & (111) & 2.908 & 2.96  &       & * \\
Au	& (100) & 1.043 & 0.86  &   1.506   & * &&     Mo	    & (100) & 2.531 & 3.18  & 2.907 & * \\
	& (110) & 1.124 & 0.83  &	        & * &&              & (110) & 2.351 & 2.80  &	    & * \\
	\midrule 
	& (111) & 1.949 & 1.92  &           & * &&      & (111) & 2.463 & 2.70   &      & *\\
Ni	& (100) & 2.041 & 2.21  &   2.380	& * &&  Ta	& (100) & 2.342 &  2.47  &2.902	& * \\
	& (110) & 2.231 & 2.29  &	        & * &&	    & (110) & 1.954 &  2.34  &	    & *\\
	\midrule
	& (111) & 1.601 & 1.34  &           & * &&     & (111) & 3.028 &    3.47  &      & *\\
Pd	& (100) & 1.728 & 1.53  &   2.003	& * &&   W & (100) & 2.967 &	3.95  &3.265 & *\\
	& (110) & 1.890 & 1.57  &	        & * &&     & (110) & 2.826 &	3.23  &       & *\\
	\midrule

	& (111) & 1.978 & 1.48 &            & *  &&          & (111) & 0.914           & 0.80    &      & * \\
Pt	& (100) & 2.384 & 1.84 &    2.489	& *  &&   Al     & (100) & 0.891 (0.883)   & 0.92    & 1.143& -0.002 \\
	& (110) & 2.150 & 1.68 &	        & * &&          & (110) & 1.011 (0.986)   & 0.98    &       &  0.001\\
	\midrule
	Mg	& (0001) & 0.377 & 0.54  & 0.760  & * &&  Zr	& (0001) & 1.278 & 1.97 & 1.909& *	\\
	\midrule
    Co	& (0001) & 2.028 & 2.11 & 2.522   & *  &&  Ti	& (0001) & 5.577 & 1.61 & 1.989& *\\
	\bottomrule
\end{tabular}
\begin{tablenotes}
\item * $\langle q \rangle < 10^{-4}$
\end{tablenotes}
\label{tab:SurfaceEnergies}

\end{table}
\end{turnpage}
\subsection{Ordered structures}
\begin{figure}
\centering
\includegraphics[width=0.5\linewidth]{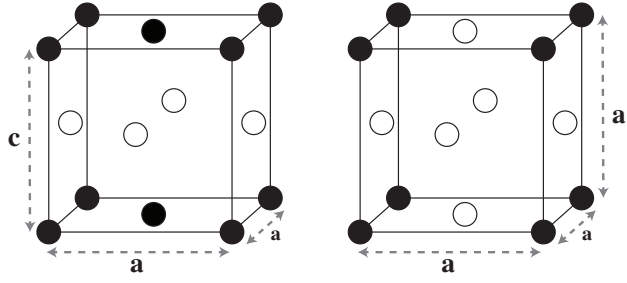}
\caption{Structures of the ordered alloys L$1_0$ (left) and L$1_2$ (right). L$1_0$ structures present alternating layers of A and B atoms, so any charge transfer between the two elements will result in modification of the $c/a$ ratio exhibited by the crystal.}
\label{fig:orderedstructures}
\end{figure}

Ordered alloys are composed of at least two elements which alternate in position in a regular pattern. Two ordered structures (L$1_2$ and L$1_0$) were used to study how DR-EAM predicts charge transfer and structural changes in alloys
(see Figure. \ref{fig:orderedstructures}).
L$1_0$ alloys have \ce{AB} composition and space group P4/mmm, which exhibits a layered structure.  The basis vectors for L$1_0$ with two components, A and B, are
\begin{eqnarray*}
    \mathbf{r}_1 &=&0\hspace{2.9cm} \textrm{(A)},\\
    \mathbf{r}_2 &=&\frac{1}{2}\,a\,\hat{x}+\frac{1}{2}\,c\,\hat{z}\hspace{1cm}\textrm{(B)},
\end{eqnarray*}
where \textit{a} and \textit{c} are the lattice parameters. The alternation of A and B layers lies along the $c$ direction, while all atoms along the two directions with the $a$ lattice parameters are identical.  This means that charge transfer between the two elements will result in modification of the $c/a$ ratio exhibited by the crystal.

L$1_2$ alloys have a \ce{AB3} composition and space group Pm$\bar{\mathrm{3}}$m. The basis vectors for L$1_2$ with two components, A and B, are
\begin{eqnarray*}
    \mathbf{r}_1&=&0\hspace{2.9cm} \textrm{(A)}~,\\
    \mathbf{r}_2&=&\frac{1}{2}\,a\,\hat{y}+\frac{1}{2}\,a\,\hat{z}\hspace{1cm}\textrm{(B)}~,\\
    \mathbf{r}_3&=&\frac{1}{2}\,a\,\hat{x}+\frac{1}{2}\,a\,\hat{z}\hspace{1cm}\textrm{(B)}~,\\
    \mathbf{r}_4&=&\frac{1}{2}\,a\,\hat{x}+\frac{1}{2}\,a\,\hat{y}\hspace{1cm}\textrm{(B)}~,
\end{eqnarray*}
where \textit{a} is the lattice parameter.  Each of the A atoms in an L$1_2$ structure is surrounded by a symmetric set of B atoms, so charge transfer between A and B will  result in isotropic volume contraction due to electrostatic interactions.

Ordered structures were formed by replacing appropriate atoms in a bulk FCC lattice of A atoms with B atoms. Following this replacement, the geometry of the periodic box was optimized, while allowing the fluctuating densities in DR-EAM to re-optimize simultaneously with each change in the geometry.

Bulk properties of the L$1_0$ and L$1_2$ ordered alloys (lattice constants, bulk modulus, shear modulus, Poisson's ratio and energy per atom) are compared with DFT calculations and experimental values (when known) in Tables \ref{tab:L12Properties} and \ref{tab:L10Properties}.
Significant charge transfer was observed between the two components of the ordered structures. 
The direction and magnitude of the transfer is governed by the electronegativities ($a_1$) of the two elements.
Even with charge transfer, for L1$_2$ structures, most of the lattice constants are nearly equal to the structures optimized under nonpolarizable EAM.   
This is likely due to cancellation of added A-B attraction with the simultaneous B-B repulsion in L$1_2$ structure. Even though the  DR-EAM and EAM structures are similar, the bulk modulus, shear modulus, Possion's ratio and energy per atom do exhibit small changes due to charge transfer. 
In the L$1_0$ structures (see Table \ref{tab:L10Properties}), the DR-EAM optimized geometries are 
different from 
structures optimized without charge transfer. However, when compared with DFT structures or experimental lattice parameters, the two EAM methods are remarkably similar.
Some notable exceptions (e.g., NiPt and AuCu) occur when charge transfer between the two metallic components is relatively large. Inter-metallic charge transfer in L$1_0$ alloys will usually increase in-layer Coulombic repulsion, while simultaneously increasing inter-layer attraction. This results in changes to the $c/a$ ratio that appear to depend largely on the extent of the charge transfer. However, very large charge transfers can also alter intermetallic pair potentials by changing the density ratios in Eq. \ref{JohnsonModified}, modifying the equilibrium geometry of the alloy.

\begin{turnpage}
\begin{table}
\setlength{\tabcolsep}{6pt}
\def\arraystretch{0.8}
\caption{The lattice constants ($a$), bulk modulus $(B)$, shear modulus $(G)$, Poisson's ratio $(\sigma)$, energy per atom $(E_a)$ 
and charge on atom B at surface $\langle q_\mathrm{B} \rangle$ for L$1_2$ (\ce{AB3}) alloys. Note: $\langle q_\mathrm{A} \rangle = - 3 \langle q_\mathrm{B} \rangle$ for these alloys. 
For each alloy, the top row is the value calculated using DR-EAM, the middle row is from nonpolarizable EAM, and the lower values are derived from plane-wave DFT calculations.\cite{mpApi}}
\centering 
\begin{tabular}{c |r r r r r r c | c|r r r r r r} 
\toprule
Alloy & $a$ (\si{\angstrom})& $B$ (GPa) & $G$ (GPa) & $\sigma$ & $E_a$ (eV)& $\langle q_\mathrm{B} \rangle$ (e) &&Alloy & $a$ (\si{\angstrom})& $B$ (GPa) & $G$ (GPa) & $\sigma$ & $E_a$ (eV) & $\langle q_\mathrm{B} \rangle$ (e) \\
     \midrule
NiAg$_3$ & 3.99 & 111.70 & 39.01 & 0.34 & -3.20 & -0.002 && PbPd$_3$ & 4.17 & 134.67 & 36.20 & 0.37 & -3.22 &  0.014\\
         & 3.99 & 111.70 & 39.01 & 0.34 & -3.20 & -     &&          & 4.17 & 134.66 & 36.19 & 0.37 & -3.21 & - \\
         & 4.00 & 64     & 27    & 0.33 & -3.37 & -     &&          & 4.12 & 132    & 47    & 0.35 & -5.11 & - \\
    
\midrule
CuAu$_3$ & 3.99 & 147.71 & 37.95 & 0.37 & -4.09 & -0.047 && AgAu$_3$ & 4.09 & 146.81 & 35.77 & 0.38 & -3.78 &  -0.043\\
         & 3.99 & 147.57 & 37.88 & 0.38 & -3.98 & -     &&          & 4.09 & 146.70 & 35.73 & 0.38 & -3.72 &  - \\
         & 4.05 & 143    & 31    & 0.41 & -3.49 & -     &&          & 4.17 & 126    & 23    & 0.44 & -3.19 &  -\\
     
\midrule
AuCu$_3$ & 3.76 & 122.06 & 50.17 & 0.31 & -3.94 & 0.057 && NiAu$_3$ & 3.98 & 153.92 & 37.58 & 0.38 & -4.19 &  -0.049 \\
         & 3.76 & 122.04 & 50.01 & 0.31 & -3.86 & -     &&          & 3.98 & 153.72 & 37.50 & 0.38 & -4.13 &  -     \\
         & 3.77 & 142    & 52    & 0.35 & -3.92 & -     &&          & 4.02 & 147    & 22    & 0.43 & -3.80 &  -      \\
     
\midrule
PdCu$_3$ & 3.73 & 132.73 & 42.50 & 0.34 & -3.61& 0.030 && PtCu$_3$ & 3.64 & 157.64 & 59.55 & 0.33 & -4.46 & 0.051\\
         & 3.72 & 132.73 & 54.65 & 0.31 & -3.79 & - &&      & 3.64 & 157.22 & 59.42 & 0.33 & -4.42 & -        \\
         & 3.71 & 146    & 53    & 0.36 & -4.47 & - &&          & 3.72 & 159    & 59    & 0.35 & -4.72 & -\\
     
\midrule
PtFe$_3$ & 3.55 & 236.48 & 94.07 & 0.32 & -5.17 & 0.054  && FePd$_3$ & 3.82 & 175.88 & 57.43 & 0.35 & -4.16 & -0.031 \\
         & 3.56 & 235.53 & 93.52 & 0.32 & -5.11 & -      &&          & 3.82 & 175.78 & 57.42 & 0.35 & -4.13 & - \\
         & 3.75 & 180    & 50    & 0.40 & -7.94 & -      &&          & 3.89 & 174    & 76    & 0.32 & -6.11 & - \\
     
\midrule
AlNi$_3$ & 3.66 & 111.60 & 48.83 & 0.30 & -4.37 & 0.014&\\
         & 3.65 & 111.56 & 48.74 & 0.30 & -4.36 & -&\\
         & 3.56 & 117    & 96    & 0.29 & -5.70 & -&\\
    
\bottomrule
\end{tabular}
\label{tab:L12Properties}
\end{table}
\end{turnpage}
\begin{turnpage}
\begin{table}[h!]
\caption{The lattice constants ($a$ and $c$), bulk modulus $(B)$, shear modulus $(G)$, Poisson's ratio $(\sigma)$, energy per atom $(E_a)$ 
and charge on B atoms $\langle q_\mathrm{B} \rangle$ for L$1_0$ (\ce{AB}) alloys. Note: $\langle q_\mathrm{A} \rangle = - \langle q_\mathrm{B} \rangle$ for these alloys. 
For each alloy, the top row is the value calculated using DR-EAM, the second row is from nonpolarizable EAM, the third row are derived from plane-wave DFT calculations in Ref. \onlinecite{mpApi} and, the bottom row are the experimental values in Ref. \onlinecite{latticeExp} (when available).  \label{tab:L10Properties}}
\setlength{\tabcolsep}{4.3pt}
\def\arraystretch{0.63}
\begin{tabular}{l | r r r r r r r r c | c | r r r r r r r r}
\toprule
Alloy & $c$ (\si{\angstrom})& $a$ (\si{\angstrom})&$c/a$& $B$ (GPa) & $G$ (GPa) & $\sigma$ & $E_a$ (eV) &$\langle q_\mathrm{B} \rangle$ (e) &&Alloy & $c$ (\si{\angstrom})& $a$ (\si{\angstrom})&$c/a$& $B$ (GPa) & $G$ (GPa) & $\sigma$ & $E_a$ (eV)&$\langle q_\mathrm{B} \rangle$ (e)\\
\midrule
AgTi & 4.26 & 4.09  & 1.04  & 117.33 & 40.74 & 0.34 & -3.95 & 0.090 && FePd & 3.62 & 3.77 & 0.95 & 179.37 & 67.84 & 0.33 & -4.27 & -0.070\\
     & 4.29 & 4.08  & 1.05  & 117.22 & 40.58 & 0.34 & -3.70 & -     &&      & 3.62 & 3.77 & 0.96 & 178.99 & 67.65 & 0.33 & -4.23 & -    \\
     & 4.17 & 4.085 & 1.02  & 111    & 37    & 0.36 & -5.43 & -     &&      & 3.78 & 3.84 & 0.98 & 163    & 82    & 0.3  & -6.88 & -    \\
     & 4.09 & 4.09  & 0.993 & -      &  -    &  -   &  -    & -     &&      & 3.72 & 3.85 & 0.96 & -      & -     & -    & -     & -    \\
\midrule
AlMg & 4.29 & 4.24 & 1.01 & 73.19 & 35.30 & 0.29 & -2.61  & 0.050 &&   FePt & 3.49 & 3.77 & 0.92 & 236.42 & 75.57 & 0.35 & -5.66 & -0.104\\
     & 4.29 & 4.24 & 1.01 & 73.47 & 35.50 & 0.29 & -2.58  & -      &&        & 3.53 & 3.75 & 0.94 & 234.42 & 75.28 & 0.35 & -5.60 & -\\
     & 4.23 & 3.02 & 1.40 & 58    & 33    & -    & -2.68  & -      &&        & 3.72 & 3.76 & 1.01 & 201    & 98    & 0.3  & -7.5  & -\\
     & -    & -    & -    & -     & -     &    - &    -   & -      &&        &3.71  & 3.83 & 0.96 & -      &-      &    - &-      & -\\
\midrule
AlTi & 4.27 & 4.10 & 1.04 & 81.09 & 29.98 & 0.33 & -4.29 & 0.020 && MgNi & 4.14 & 3.99 & 1.03 & 101.53 & 38.89  & 0.33 & -2.54 & -0.079 \\
     & 4.27 & 4.10 & 1.04 & 81.17 & 29.91 & 0.33 & -4.28 & -     &&      & 4.16 & 3.99 & 1.04 & 101.12 & 38.63  & 0.33 & -2.47 & -     \\
     & 4.08 & 3.99 & 1.02 & 115   & 76    & 0.25 & -6.22 & -     &&      & 3.16 & 2.99 & 1.05 & 150    & 55     & 0.35 & -3.84 & -     \\
     & 4.06 & 3.99 & 1.02 &   -    &    -   &-      & -  & -     &&      & -    & -    & -    & -      &-       &-     &-      & -     \\

\midrule
AuCu & 3.52 & 4.04 & 0.87 & 136.21 & 43.23 & 0.35 & -4.06 & 0.119 && MgZr & 4.85 & 4.22 & 1.14 & 105.69 & 41.26 & 0.32 & -4.03 & -0.004 \\
     & 3.66 & 3.97 & 0.92 & 135.45 & 42.84 & 0.35 & -3.95 & -     &&      & 4.85 & 4.22 & 1.14 & 105.69 & 41.26 & 0.32 & -4.03 & -     \\
     & 3.66 & 4.05 & 0.90 & 135    & 37    & 0.4  & -3.73 & -     &&      & 4.46 & 3.16 & 1.41 & -      & -     & -    & -4.99 & -     \\
     & 3.66 & 3.92 & 0.92 & -      & -     & -    &-      & -     &&      & -    & -    &     -&-       & -     &-     & -     & -     \\
\midrule
CoPt & 3.62 & 3.77 & 0.95 & 213.43 & 61.59 & 0.36 & -5.52 & -0.103 && NiPt & 3.59 & 3.77 & 0.95 & 202.64 & 59.68 & 0.36 & -5.56 & -0.094\\
     & 3.64 & 3.77 & 0.96 & 212.85 & 64.82 & 0.36 & -5.46 & - &&          & 3.43 & 3.86 & 0.88 & 200.37 & 58.23 & 0.36 & -5.51 & -\\
     & 3.72 & 3.82 & 0.97 & 216    & 114   & 0.29 & -6.67 & - &&          & 3.63 & 3.85 & 0.94 & 214    & 95    & 0.31 & -6.02 & -\\
     & 3.70 & 3.81 & 0.97 &  -     & -     & -    & -     & - &&          & 3.58 & 3.81 & 0.93 & -      & -     & -    & -     & -\\

\midrule
FeNi & 3.51  & 3.51 & 1.00 & 162.43 & 83.79 & 0.27 & -4.51 & -0.017 && PdTi & 3.97 & 4.09 & 0.97 & 143.42 & 45.36 & 0.35 & -4.65 & 0.119\\
     & 3.53 & 3.51 & 1.00 & 162.39 & 83.76 & 0.27 & -4.51 & -     &&      & 4.03 & 4.06 & 0.99 & 142.77 & 44.99 & 0.35 & -4.54 & -\\
     & 3.58 & 3.55 & 1.00 & 187    & 103   & 0.29 & -7.19 & -     &&      & 3.87 & 2.84 & 1.46 &      - & -     & -    & -7.07 & -\\
     & 3.61 & 3.56 & 0.98 &  -     & -     & -    & -     & -     &&      &-     &-     &-     &-       &-      & -    &-      & -\\
   \bottomrule
\end{tabular}
\end{table}
\end{turnpage}

\subsection{Metal slabs in a uniform field}
An atomic partial charge $q_i$ in an electric field $\mathbf{E}$ feels a physical force $\left( \mathbf{F}_i = e q_i \mathbf{E} \right)$ due to that field. To derive forces on the \textit{fluctuating charge} variables due to the presence of an external field, we define a potential that depends on the location of each atom,
\begin{equation}
    V_\mathrm{ext} = - \sum_{i=1}^{N}\mathrm{e} q_i \mathbf{r}_i \cdot \mathbf{E} .
\end{equation}
This potential provides a fluctuating charge force due to the field,
\begin{equation}
    - \frac{\partial V_\mathrm{ext}}{\partial q_i} = \mathrm{e} \mathbf{r}_i \cdot \mathbf{E},
\end{equation}
which depends on the position of the atom relative to the origin of the coordinate system.  (A spatial derivative of $V_\mathrm{ext}$ also yields the correct physical force on the atom's coordinates.)


We note that this uniform field is not realizable in any experiment, particularly when discussing a conducting slab. Uniform fields may be approximated in a region between charged plates, but interior to a conductor, the field will be shielded by the skin, and an external field should not interact with the atoms on the interior.  However, it is useful to test the behavior of the new methodology on exposure to the field. If DR-EAM can approximate the correct behavior of a conductor, the response should be an effective cancellation of the external field inside a metal slab.

A local electric dipole density can be defined between consecutive atomic layers. The origin of a local coordinate system ($O_n$) is set between layers $n$ and $n+1$, and the local dipole density  between these layers is measured with respect to this origin,
\begin{equation}
\mathbf{P}_n = \frac{\mathrm{e}\langle q_{n} \rangle \mathbf{R}_n +e\langle q_{n+1} \rangle \mathbf{R}_{n+1}}{v},
\end{equation}
where $\langle q_n \rangle$ is the average charge in the n$^\mathrm{th}$ layer, $\mathbf{R}_n$ and $\mathbf{R}_{n+1}$ are the average positions of the two layers with respect to $O_n$, and $v$ is the volume between the two layers $(v = L_x L_y \Delta z = A \Delta z)$.

To test the shielding properties of DR-EAM, an electric field was applied along the $z$-axis of a  slab comprising 18 atomic layers of each metal. Because the local origin is located half-way between each layer, the polarization density,
\begin{equation}
\mathbf{P}_{n}=\mathrm{e}\frac{\langle q_{n+1} \rangle - \langle q_n \rangle }{ 2 A } \hat{z}~,
\end{equation}
depends only on the average charges on the two layers.  The bound charge density inside the slab, $\rho_b = - \nabla \cdot \mathbf{P}$. Since our external field (and polarization density) both point along the $z$-axis, we can simply write the bound charge density, $\rho_b = - \partial P / \partial z$. 

\begin{figure}[H]
\centering

  \includegraphics[width=0.85\linewidth]{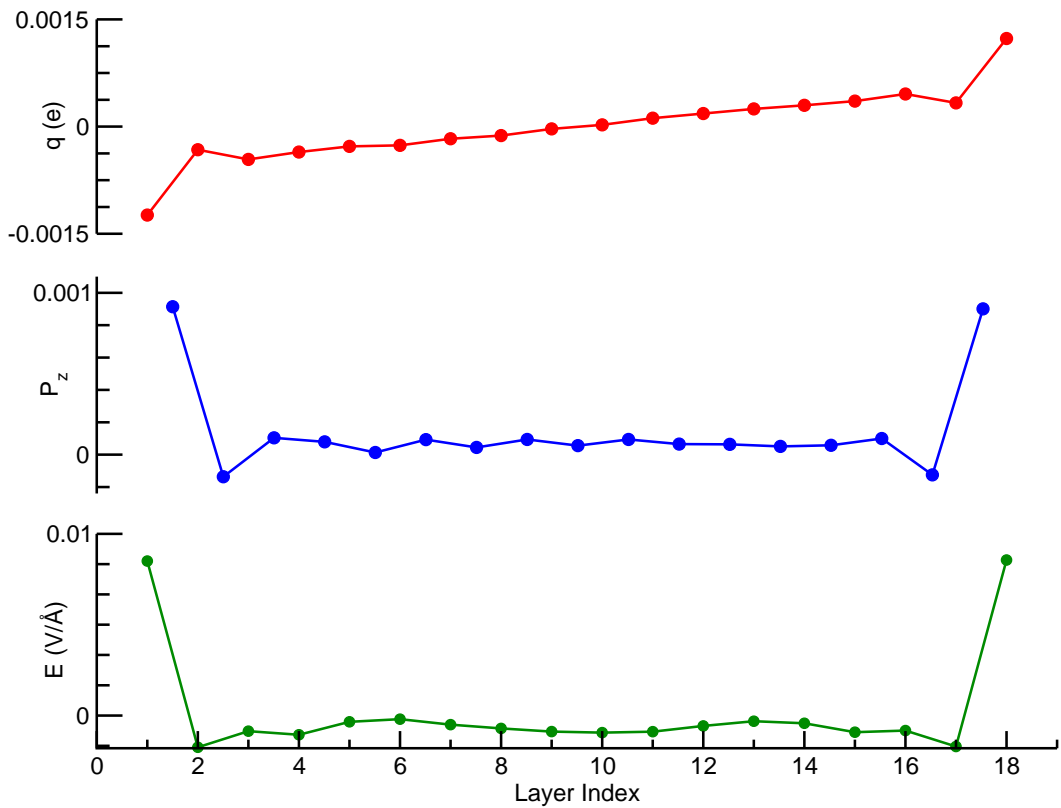}
  \caption{The response of an 18-layer DR-EAM Platinum (111) slab exposed to an external uniform electric field (0.01~V/\AA\ along the $z$-axis). The average charge in each layer (top) displays a nearly linear dependence on the $z$ coordinate of the layer. The dipole density(middle), in units of $\frac{\mathrm{e}}{2A}$, is approximately constant in the interior of the slab. The net electric field (bottom) exhibits a nearly complete screening in the interior of the slab.  Only the outermost atomic layers feel the full external field, while the response of the DR-EAM densities effectively screens the interior.  The model has an effective penetration depth of 2-3 atomic layers before complete screening is recovered. }
  
\label{fig:screen}
\end{figure}

When an electric field is applied to a Pt slab with 18 layers, the resulting charge distribution in DR-EAM screens the electric field in the interior of the slab. 
The average charges in each layer are approximately linear in $z$ in the interior of the slab, which yields a nearly constant dipole density in the interior, and nearly zero bound charge density inside the slab. 
Figure. \ref{fig:screen} shows the average charge in each layer as well as the local dipole density and screened electric field values at the surface and interior of the slab.
These are the expected responses of a classical conductor in an external field.  The electric field penetrates up to three layers deep from the atomic skin, but is fully screened beyond that.

\subsection{Image charge effects}
When a ion is placed directly above a conducting metal slab, the ion polarizes the slab and an oppositely charged image is induced on the surface of the slab directly beneath the ion.
Classical treatments of charges at the surfaces of infinite planar conductors predict an effective interaction of the ion with its own image, $U_\mathrm{eff}(d) = - q^2 \mathrm{e}^2/(4 \pi \epsilon_0 \, 2 d)$, where $d$ is the separation of the ion from the surface of the 
conductor, and $q$ is the 
charge on the ion (in units of electrons, $e$).

\begin{figure}[H]
\centering
  \includegraphics[width=0.7\linewidth]{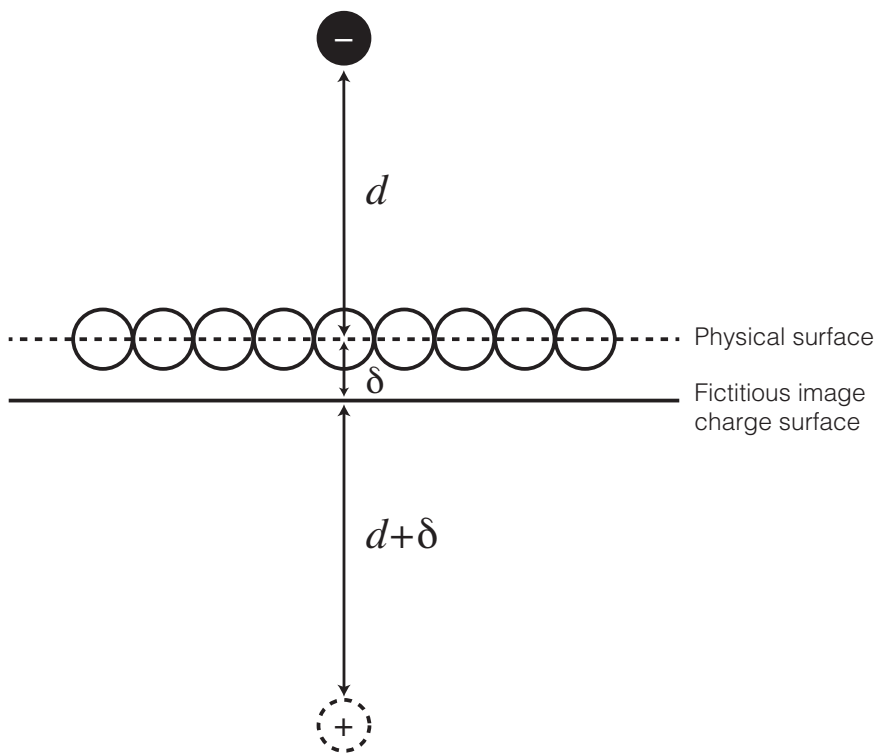}
  \caption{Configurational energies as a function of the distance $(d)$ of a point charge from the surface are fit to find an offset $\delta$ of the image charge surface from the top layer of atoms, and a scaling parameter $(s \sim 2)$ to understand how closely DR-EAM captures classical image effects.}
\label{fig:imageScheme}
\end{figure}

When computing electrostatic interactions between atoms in the DR-EAM model and external fixed charges, we have used the Damped Shifted Force (DSF) potential. The electrostatic kernel in DSF is given by Eq. (\ref{eq:DSFPot}), so it is reasonable to fit the effective potential between the unit point charge ($q = \pm 1$) and the polarizable DR-EAM surface with a similarly damped interaction model,
\begin{equation}
    U_\mathrm{eff}(d)=-\frac{\mathrm{e}^2}{4\pi\epsilon_0}\,\left[\frac{\mathrm{erfc}\left(\alpha~s (d+\delta)\right)}{s (d + \delta) }\right]~,
    \label{eq:image}
\end{equation}
where $\alpha$ is the value of the damping coefficient (in \si{\angstrom}$^{-1}$) used in the calculation. If the DR-EAM model captures the physics of classical image charges, the fitting parameter $s$ would be expected to be exactly two for infinite planar conducting surfaces, $\delta$ measures the offset of the surface of zero potential from the atomic layer at the top of the slab 
(see Figure. \ref{fig:imageScheme}).
For a perfectly flat classical conductor, we would expect $\delta = 0$. 

\begin{figure}[H]
\centering

  \includegraphics[width=0.8\linewidth]{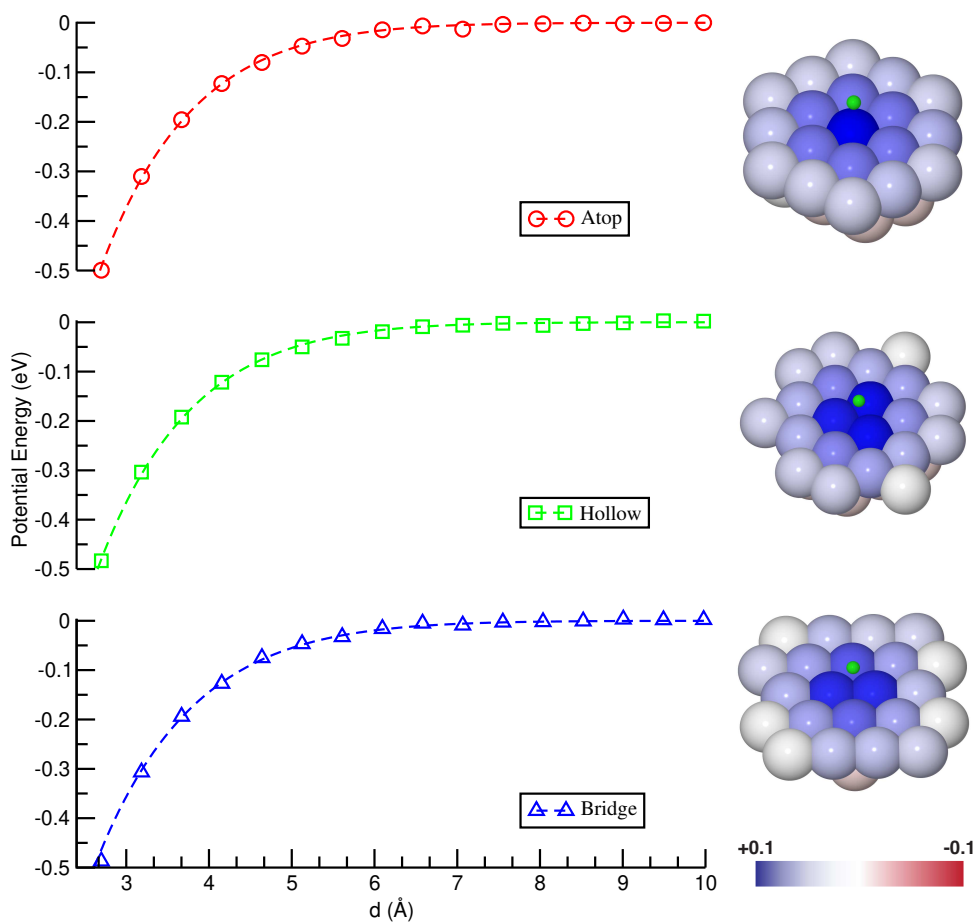}
  \caption{Effective interaction potential [Eq. (\ref{eq:image})] for a point charge model of a Chloride ion $(q = -1)$ approaching a Copper (111) surface (simulated using DR-EAM) above three different binding sites (atop, three-fold hollow, and bridge). On the right, we show the induced density changes in the metal (represented by changes in the atomic partial charges).  Parameters for the fits of the effective interaction potential (dashed lines) are given in Table \ref{tab:imagefit} }
  
\label{fig:image}
\end{figure}

DR-EAM does exhibit image charge effects (shown graphically in Figure. \ref{fig:image}).  We placed a point charge models of chloride $(q=-1)$ ions above a range of metal surfaces, including Copper, Gold, and Platinum (111) and (100) surfaces.  The ions are brought close to the surface above common sites (atop, hollow, and bridge) on these surfaces and the partial charges are allowed to find their lowest energy configurations. The surfaces polarize directly beneath the probe charge, and the distance dependence of the total interaction is used to fit $d$, $s$, and $\delta$ values in Eq. (\ref{eq:image}). For all of the surfaces presented in Table \ref{tab:imagefit}, we note that the scaling approaches the classical conductor value $s \rightarrow 2$, and $\delta$ sits approximately 0.5 layers beneath the atomic coordinates of the surface atoms.

\begin{table}[H]
\caption{Parameters of the effective image charge potential [Eq. (\ref{eq:image})], $s$ and $\delta$ (\AA), for three different metals (Cu, Au, Pt) with exposed (100) and (111) facets.  A point charge representing a chloride ion is brought close to the surface, and the effective potential is computed.  For a (flat) infinite planar conductor, we would expect $s \rightarrow 2$ and $\delta \rightarrow 0$.  This data suggests that the atomistic polarizable model in DR-EAM is capturing most of the classical image effect.} 
\setlength{\tabcolsep}{6pt}
\def\arraystretch{0.75}
\centering 
\begin{tabular}{l|c c c c c c c} 
\toprule 
Metals&Facet&\multicolumn{2}{c}{Atop}&\multicolumn{2}{c}{Hollow}&\multicolumn{2}{c}{Bridge}\\
\midrule
&&$s$&$\delta$&$s$&$\delta$&$s$&$\delta$\\
\midrule
Copper&100&1.86&0.84&1.76&1.08&1.80&0.98\\
&  111  &1.84&0.72&1.80&0.80&1.74&0.98\\
\midrule
Gold&100&1.78&1.25&1.78&1.24&1.70&1.49\\
&111&1.84&1.07&1.79&1.22&1.79&1.22\\
\midrule
Platinum&100&2.19&0.36&1.51&1.69&1.64&1.31\\
&111&2.18 &0.34 &1.59 &1.36 &1.62 &1.26 \\
\bottomrule
\end{tabular}
\label{tab:imagefit}
\end{table}

\section{CONCLUSIONS}

DR-EAM is a polarizable force field for metals where each metal atom has an additional variable, or partial charge, that represents a fluctuation in the local valence density on that atom. 
The dynamics of the partial charges are calculated using an extended Lagrangian, and can aid in simulating polarization and charge transfer effects in systems that contain these metals. 
The polarization catastrophe in traditional electronegativity equalization models is solved using a sixth order polynomial for the self potential. 
The coefficients of this polynomial are tied to thermodynamically- and experimentally-derived data, notably the Pauling-Allred electronegativities, Hall coefficients, higher ionization energies, and bulk polarizabilities of these metals. 

A number of important physical properties of metals are captured by DR-EAM. The partial charges distribute on the surface in response to external fields and produce screening of the electric field inside a metal slab. 
In addition, the surface charge response reproduces the classical image charge effect for point charges. 
These effects cannot be modeled using nonpolarizable versions of the embedded atom method.  
Additionally, all of the strengths of EAM (i.e. reasonably quantitative results for bulk elastic constants and surface energies) have been retained in DR-EAM.  
Relative to an unmodified EAM simulation, the extended Lagrangian in DR-EAM adds an additional 12.5\% to the computational cost associated with a 2000 atom, 1~ns simulation.  
This is a relatively small additional cost given the new capability to simulate interactions with charged and polar molecules adsorbed on metal surfaces.
\begin{acknowledgments}
  Support for this project was provided by the National Science
  Foundation under Grant No. CHE-1663773. Computational time was provided
  by the Center for Research Computing (CRC) at the University of
  Notre Dame.
\end{acknowledgments}
\bibliography{fdeam}
\end{document}


\title{Supplemental Materials: Polarizable Potentials For Metals: The Density Readjusting Embedded Atom Method (DR-EAM)}

\author{Hemanta Bhattarai}
\affiliation{Department of Physics, \\
  University of Notre Dame, Notre Dame, IN 46556}
\author{Kathie E. Newman}
\affiliation{Department of Physics, \\
  University of Notre Dame, Notre Dame, IN 46556}
\author{J. Daniel Gezelter}
\email{gezelter@nd.edu}
\affiliation{251 Nieuwland Science Hall, \\
  Department of Chemistry \& Biochemistry, \\
  University of Notre Dame, Notre Dame, IN 46556}

\date{\today}

\begin{abstract}
  This document includes relations used to determine the bulk modulus, shear modulus and Poisson's ratio. Additional experimental data used in DR-EAM to fit the self potential is presented along with the definitions used during fitting. A plot of the self potentials for partial charges in a reduced range (-0.4e to 0.4e) is also included. 
\end{abstract}

\maketitle

\section{Bulk Metals}
For sufficiently small deformations, the Lagrangian stress ($\tau$) and strain ($\eta$) are related by generalized Hooke's law
\begin{equation}
    \tau_{\alpha\beta} = \sum_{\gamma,\delta=\{x,y,z\}} c_{\alpha\beta\gamma\delta} \,\eta_{\gamma\delta}
\end{equation}
where the coefficients $c_{\alpha\beta\gamma\delta}$ are components of the fourth-rank stiffness tensor. This can be expressed in a more condensed form,
\begin{equation}
    \tau_i=\sum_{j=1}^6 C_\mathrm{ij} \eta_j
\end{equation}
These tensors are in Voigt notation. Values for the two indices $(i, j)$ from 1--6 denote locations in the rank 2 tensors, $1\mapsto xx, 2\mapsto  yy, 3\mapsto zz, 4 \mapsto yz, 5 \mapsto xz, 6 \mapsto xy$.


The bulk modulus, $B$, was computed using the Voigt average,\cite{Jong:2015bs}
\begin{equation}
    9 B =\left(C_{11}+C_{22}+C_{33}\right) +2\left(C_{12}+C_{23}+C_{31}\right)
\label{eq:BM}
\end{equation}
The shear modulus, $G$, was also computed using the Voigt average,
\begin{equation}
    15 G =\left(C_{11}+C_{22}+C_{33}\right) - \left(C_{12}+C_{23} + C_{31}\right) +3\left(C_{44}+C_{55}+C_{66}\right) 
\label{eq:SM}
\end{equation}
Poisson's ratio $(\sigma)$ is a dimensionless number describing the lateral response to loading. We used the bulk and shear moduli computed above to calculate Poisson's ratio.
\begin{equation}
\sigma = \frac{3\,B - 2\,G}{6\,B+2\,G}    
\end{equation}

\section{Surface Energies}
Surface energies were computed by slicing a crystal along a facet perpendicular to the given $(hkl)$ Miller indices, and rotating a large block of this surface normal to the $z$-axis of the simulation cell.  Surface energies are computed by comparing the difference in energy between the bulk (periodic) system with the same number of atoms and a system with a vacuum layer that is larger than any cutoff radius used in the system.  Each of these vaccuum-gapped simulations exposes two identical surfaces with total area $A = 2 L_x L_y$.  No reconstruction or relaxation is allowed before the energy calculation.

\section{Experimental Energies Used in Fitting the Model}
Although reduced-partial charges are used to represent different ionic states in DR-EAM, the energies used to fit the self energy are obtained directly from gas-phase experiemental data on ionization potentials and electron affinities.
It is helpful to define the energies that are being used for the fits.  For ionization potentials, the definitions are straightforward,
\begin{align}
\mathrm{M} \rightarrow \mathrm{M}^+ + \mathrm{e}^- ~~~~~~~ & \Delta E = \mathrm{IP}_1 \label{eq:IP1} \\
\mathrm{M}^+ \rightarrow \mathrm{M}^{2+} + \mathrm{e}^- ~~~~~~~ & \Delta E = \mathrm{IP}_2 \label{eq:IP2} \\
\mathrm{M}^{2+} \rightarrow \mathrm{M}^{3+} + \mathrm{e}^- ~~~~~~~ & \Delta E = \mathrm{IP}_3 \label{eq:IP3} \\
\mathrm{M}^{3+} \rightarrow \mathrm{M}^{4+} + \mathrm{e}^- ~~~~~~~ & \Delta E = \mathrm{IP}_4 \label{eq:IP4} \\
\mathrm{M}^{4+} \rightarrow \mathrm{M}^{5+} + \mathrm{e}^- ~~~~~~~ & \Delta E = \mathrm{IP}_5 \label{eq:IP5} 
\end{align}
where the various IP$_n$ values are obtained from gas phase ionization potentials.  

Electron affinities (where known) are defined by,
\begin{equation}
\mathrm{M} + \mathrm{e}^- \rightarrow \mathrm{M}^-  ~~~~~~~  \Delta E = -\mathrm{EA}_1
\label{eq:EA1}
\end{equation}
(note the negative sign) and are typically obtained by negative ion photodetachment energies, \textit{i.e.} the energy to reverse this reaction.\cite{Hotop:1973} In some cases, notably for \ce{Mg}, the reported electron affinities are predictions based on condensed phase stabilities of various ions.\cite{Mg:EA} 

\begin{table}
\caption{Experimental electron affinities (see Eq. \ref{eq:EA1})~\cite{CuAgMo:EA, Au:EA, NiPd:EA, Co:EA, Pt:EA, Al:EA, Pb:EA, Fe:EA, Ta:EA, W:EA, Mg:EA, Ti:EA, Zr:EA} and ionization potentials (see Eqs. \ref{eq:IP1}-\ref{eq:IP5}) for the elements parameterized in this work. For most of the metals, only the first four ionization energies were used in the fits to the self energies.  All energies are given in eV.}
\centering
\begin{tabular}{l r r r r r r r}
\toprule
Element & -EA$_1$ & IP$_1$ & IP$_2$ & IP$_3$ & IP$_4$&IP$_5$\\
\midrule
Cu   &   -1.23  & 7.72 & 20.29 & 36.84  &57.37&-\\
Ag   &   -1.30  & 7.57 &21.48  &34.83  & 49.00&-\\
Au   &   -2.31  & 9.23 & 20.20 & 29.95& 44.98&-\\
Ni   &   -1.16  & 7.64 & 18.17 & 35.19 & 54.92&-\\
Pd   &   -0.56 & 8.33 & 19.43 & 32.93& 45.99&-\\
Pt   &   -2.13 & 8.95 & 18.56 & 29.02 & 43.01&-\\
Al   &   -0.43 & 5.98 & 18.83 & 28.44&119.99&-\\
Pb   &   -0.36 & 7.41 & 15.03 & 31.93 & 42.33&68.81\\
Fe   &   -0.15 & 7.90 & 16.19 & 30.65 & 54.91&-\\
Mo   &   -0.74 &7.09 & 16.16 & 28.26 & 40.32&-\\
Ta   &   -0.32 & 7.55 & 16.17 &23.11  & 35.03&-\\
W   &   -0.81 & 7.86 & 16.37 & 26.01 & 38.24&-\\
Mg   &   0.42 & 7.65 & 15.03 & 80.14 &109.26 &-\\
Co   &   -0.66 & 7.88 & 17.08 & 33.50 &51.27 &79.50\\
Ti   &   -0.08 & 6.82  & 13.57 & 27.49&43.27&-\\
Zr   &   -0.42  & 6.63 & 13.13 & 23.17 & 34.42&80.34\\
\bottomrule
\end{tabular}
\label{table:affinityIonization}
\end{table}

\section{Self Potential }
\begin{figure}[H]
\centering
\includegraphics[width=0.85\linewidth]{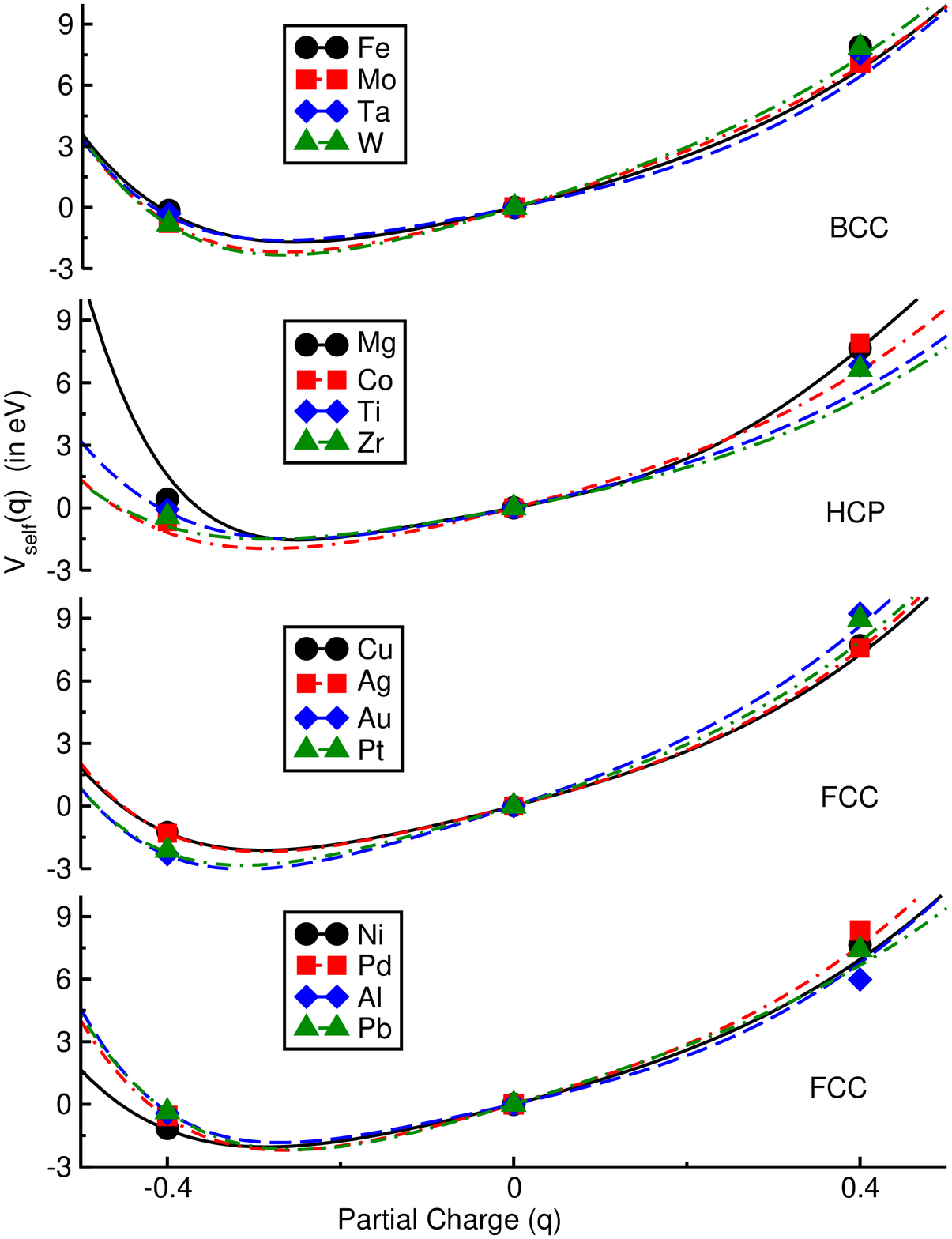}
\caption{DR-EAM self potential, $V_\mathrm{self}(q)$, for FCC, BCC, and HCP metals in the range of partial charges from -0.4$\,<q<\,$0.4.  Note that the $a_1$ and $a_2$ parameters of the $V_\mathrm{self}$ polynomial are set by Pauling electronegativity and Slater hardness, respectively, and not by IP and EA values (as in other fluctuating charge models). \label{fig:zoomedSelf}}
\end{figure}

\newpage
\bibliography{fdeam}